\documentclass[review]{elsarticle}
\usepackage{graphicx}
\usepackage{url,hyperref}
\begin{document}

\begin{frontmatter}

\title{Visceral theory of sleep and origins of mental disorders} 

\author[TSU]{Mariam M.~Morchiladze}
\ead{Marmorchila@gmail.com}
\author[TSU]{Tamila K.~Silagadze}
\ead{tamila1705@rambler.ru}
\author[NSU]{Zurab K.~Silagadze}
\ead{silagadze@inp.nsk.su}
\address[TSU]{Tbilisi State Medical University, Tbilisi, Georgia}
\address[NSU]{Novosibirsk State University  and Budker Institute of Nuclear
Physics, 630 090, Novosibirsk, Russia.}

\begin{abstract}
Visceral theory of sleep states that the same brain neurons, which process 
external information in wakefulness, during sleep switch to the processing 
of internal information coming from various visceral systems. Here we 
hypothesize that a failure in the commutation of exteroceptive and 
interoceptive information flows in the brain can manifest itself as a mental 
illness.
\end{abstract}


\end{frontmatter}

\section{Introduction}
Bipolar disorder is a nasty mental illness characterized by alternating 
episodes of mania, hypo-mania and depression, intertwined with periods of 
remission \citep{1,2,3}. 

Bipolar disorder affects more than one percent of the world's population
irrespective of nationality, ethnic origin, or social status, which is
similar to that of schizophrenia --- another major psychiatric disorder
\citep{1,2}.

At present there is no generally accepted theory of the origin of mental 
disorders, and what exactly causes the illness
is still unknown. The modern neurobiological evidence of  functional and 
structural abnormalities, as well as gray-matter and white-matter changes 
related to the bipolar illness strengthen the view what seems to be clear from 
the very beginning: the bipolar disorder is a condition in which emotions gain 
too much power over behavior \citep{4,5}. At that it is impossible to indicate 
a single genetic or neurobiological cause of the disease. Diverse biological
factors can lead to abnormalities in signal transduction pathways and, as 
a result, to dysfunction of interconnected brain networks, which reveals 
itself as bipolar disorder. It seems two interrelated prefrontal-limbic 
networks play a crucial role in the pathophysiology of bipolar illness 
\citep{4}. The first network, commonly referred to as the Automatic/Internal 
emotional regulatory network, modulates amygdala responses to endogenously 
generated feeling states, induced by memories of past. The second network, 
commonly referred to as the Volitional/External regulatory network, modulates
voluntary regulation of externally induced emotional states and suppresses 
maladaptive affects. However it is not altogether clear why these 
interconnected brain networks become deteriorated and desynchronized during 
development of the illness.

Notably obsessive-compulsive symptoms, such as picking at skin (excoriation), 
are common in patients with bipolar disorders \citep{6}. There
is continuing debate about whether obsessive-compulsive disorder and bipolar 
disorder are just comorbid two independent conditions or the first is a 
specific subtype of the second \citep{6}. Sometimes the association between 
the two conditions is so persistent and striking that we have a growing 
feeling that if we understand the cause of this psychosomatic syndrome we 
understand the cause of the bipolar disorder.

There are indications that inflammation is increased in the periphery of the 
body during both manic and depressive episodes of the bipolar disorder with 
some return to normality during remissions \citep{4,7}. It seems dysregulation 
of glial-neuronal interactions, and as a result over-activity of microglia --- 
the brain's primary immune elements, is a primary source of the inflammation 
\citep{4,7}.

We see some similarity in irrational inflammatory reaction on, for example, 
spurious midge bites and abnormal emotions without apparent external cause 
characteristic in bipolar disorder: in both cases the reaction of the organism 
seems irrational. Our hypothesis is that these irrational responses are caused 
by some internal (visceral) information which is misinterpreted by the brain 
as coming from external sources. 

Emotional experience is a result of integrating by brain internal and external
information streams, the first coming from the internal bodily environment and
the second --- from sensory inputs from the outside world \citep{8,9}. Recent 
study \citep{10} hints that brain uses the notorious  principle ``divide and 
conquer'' to avoid sensory overload when presented with competing signals. 
It is not excluded that the ubiquitous presence of sleep or sleep-like states
in all animals with brains, from worms to humans, indicates that the same 
principle underlines the brain's information-processing architecture that 
copes with the internal and external information streams.
 
\section{The visceral theory of sleep}
Sleep is an universal and  fundamental biological phenomenon in animal kingdom.
However its genuine  function remains one of the most persistent, enigmatic  
and perplexing mysteries in biology \citep{11}. If sleep indeed plays a vital 
role, as every piece of evidence does indicate, a better understanding of 
its purpose is desired and such an understanding promises a great advance in
biology and medicine.
 
The sleep-research pioneer Allen Rechtschaffen once noted that ``if sleep does 
not serve an absolute vital function, then it is the biggest mistake the 
evolutionary process ever made'' \citep{12}. Although it was provocatively
argued that sleep is indeed an evolutionary junkyard of non-adaptation and 
essentially has no other function than simple rest when animals have nothing 
else to do \citep{13}, such an attitude is unconvincing and does not withstand 
a close inspection \citep{14}. There exist ample  evidence for a functional 
role of sleep other than inactivity, indicating that sleep must serve 
a function so important that it outweighs evolutionary pressure due to the 
inherent danger that animals assume by sleeping \citep{14}.

The visceral theory of sleep, developed by  I.~N.~Pigarev \citep{15,16,17,18}, 
identifies the longed searched vital function of sleep:  the same cortical 
neurons, which process exteroceptive information in wakefulness, during sleep
switch to the processing of information coming from the various visceral 
systems. Main proposals of the visceral theory of sleep were confirmed in 
many non-trivial experiments. The description of these experiments can be found
in the above cited literature.

Although at first sight this hypothesis seems paradoxical, it is in harmony
with findings of computational science that the universal processors are much
more effective way to build a computer than development of processors 
specialized for a single function.

\begin{figure}[htp!]
\begin{center}
\includegraphics[width=0.85\textwidth]{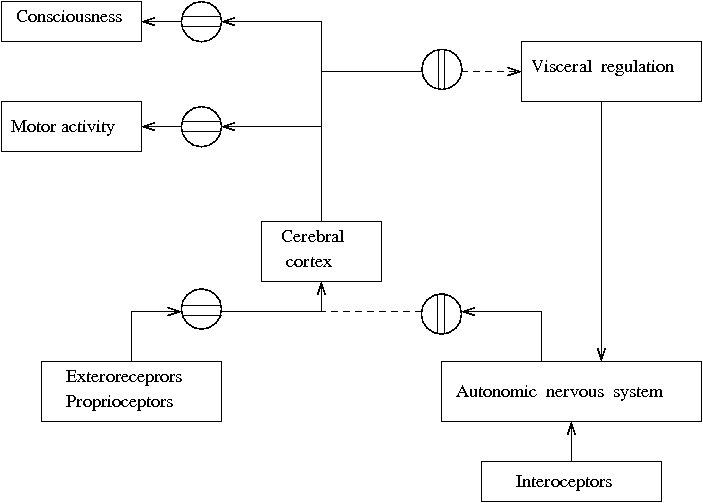}
\noindent\makebox[\linewidth]{\rule{0.6\paperwidth}{0.5pt}}
\noindent\makebox[\linewidth]{\rule{0.6\paperwidth}{0.0pt}}
\includegraphics[width=0.85\textwidth]{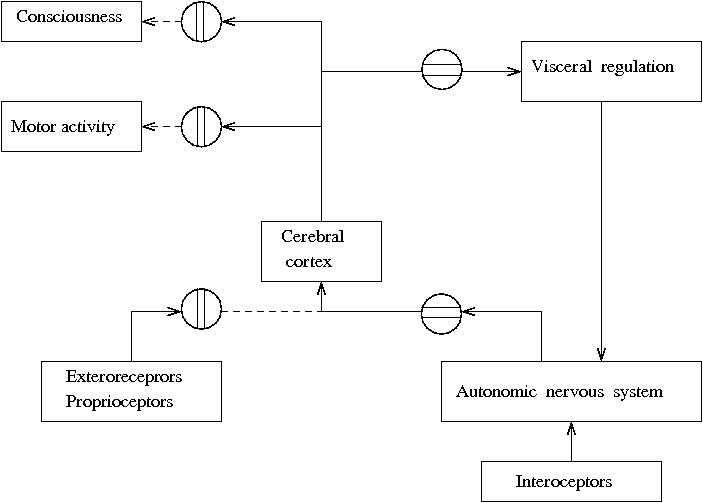}
\end{center}
\caption{Schematic view of information flow in the brain during wakefulness 
(up) and during sleep (down) as it was proposed in the visceral theory of 
sleep. Solid and dashed lines show active and blocked signal transmission 
channels, respectively. For a more detailed description, see \citep{16}.}
\label{fig1}
\end{figure}

In visceral theory of sleep cerebral cortex plays a role of universal 
processor. Information streams in the brain during wakefulness and during
sleep are schematically shown in Fig.~\ref{fig1}.
Of course Fig.~\ref{fig1} conveys only the principal idea of the visceral 
theory of sleep that the processing of  exteroceptive and interoceptive 
information streams are largely separated in time. The real architecture of  
gating elements that ensure such a separation is expected to be much more 
complex than shown in Fig.~\ref{fig1}. For example, it is a well known fact 
that in conditions when it is necessary to remain awake in spite of high 
natural sleep pressure a local sleep develops in high order cortical areas.
At that simple behavioral activities, which do not need engagement of the 
highest cortical resources, are still maintained even when part of the brain 
is sleeping \citep{19}. It is evident that to ensure such a brain plasticity
in information processing more complex gating system is needed than shown in 
Fig.~\ref{fig1}. Nevertheless even this simplified picture does allow to
explain a whole series of pathological phenomena related to the sleep-wake
cycle that are caused by desynchronization of the switching of information 
streams in the brain, as explained in detail in \citep{16}. Here we repeat
only some of these explanations to prepare a stage for our main idea. 

Dreams constitute the most frequent and harmless pathology of sleep 
\citep{16}. According
to the visceral theory of sleep, dreams happen during a brief transient period
from wake to sleep or vice verse. For example, while going to sleep, if the 
gating to the consciousness block is not timely closed and remains slightly 
ajar for a while, signals of visceral information erroneously enter the 
consciousness block. In most cases such signals will be perceived merely as 
a noise and rejected. But sometimes visceral signals can resemble archetypes 
occupying the consciousness during a wakefulness and excite the corresponding 
neurons leading to the fantastic images of a dream. 

Interestingly such a mechanism provides a ready physiological basis for 
psychoanalytic approaches based on interpretation of dreams \citep{16}. Indeed,
on the one hand the subject of dreams will most probably be determined by 
neurons having the lowest thresholds and such neurons correspond to the most 
active archetypes in the consciousness during wakefulness. On the other hand 
it seems  visceral afferent impulses constitute one of the most important 
ingredients determining unconscious psychical functions \citep{20}.

Less harmless situation can arise if the gating of the consciousness block 
does not properly switch even after long periods of time. In this case  
persistent nocturnal nightmares might constantly interrupt normal sleep. 

If upon a transition from wakefulness to sleep the consciousness block 
switch\-es normally but the switching of the motor activity block is delayed,
some visceral inputs can activate spinal cord motoneurons and lead to sudden 
limb movements. Such a sensorimotor disorder is called restless legs syndrome
and it affects from 1 to 10 percent of the population \citep{21}. 
Interestingly, it was found that restless legs syndrome prevails in  
patients with intestinal disorders \citep{22}. Visceral theory of sleep 
provides a simple mechanism, just described, underlying these, otherwise 
mysterious, comorbidities.

Another harmless sleep pathology is hypnagogic hallucinations which appear
before falling asleep under the obligatory condition of low level of 
illumination. Gates to the consciousness and motor activity blocks are still
not closed, as well as the conduction of visual information to the cortex,
but the pressure of sleep already opens the pathways for visceral 
afferentation to the cortex. Because of low level of illumination, the 
intensity of the visual stream is comparable with that of the opened visceral 
stream and visceral signals can induce visual phantoms superimposed on the 
real visual scene which is still being perceived \citep{16}. As a result 
a hallucination of large moving beetles, sensation of the wavelike movements 
of the floor or deformation of particular objects can appear. Hypnagogic 
hallucinations are eliminated if one turns the light on because the arousing
effect of bright light forces switching off of the visceral inputs. 

Visceral theory of sleep can simply explain even the mysterious phenomenon of
somnambulism (sleepwalking) \citep{16}, which in the past was thought to be 
due to demons or other supernatural phenomena \citep{23,24}. Sleepwalking is 
more common in childhood, where it is typically benign and as a rule disappears
by late adolescence, than in adults, where it has substantial harm potential. 
It occurs when the brain does not fully awaken from a deep sleep. In 
particular, according to the visceral theory of sleep, the gate to the 
consciousness block for some reason remains closed after arousal. 
Sleepwalking can last from less than a minute up to an hour or more. Sometimes 
sleepwalkers can perform very complex things during this time interval, like 
prepare a meal, play a musical instrument, drive a car, vocalize or even have 
conversations or sex with strangers, yet they does not have conscious 
awareness and their responsiveness to the environment, as well as their mental 
abilities such as memory, planning, and interacting with others may be greatly 
reduced or entirely lacking \citep{24}.

If during a sleep the gates to the consciousness and motor activity blocks
become simultaneously opened, a condition more severe than the restless legs 
syndrome, the so-called REM behavior disorder can occur. This disorder was 
first described in \citep{25}. In REM behavior disorder unusually vivid 
dreams and nightmares are accompanied by vigorous behavior and violent 
confrontations with the phantoms of dreams. For example,  a 73-year-old very 
pleasant, mild, and considerate retired librarian during his  vivid and 
violent dreams used to jump out from the bed, breaking objects in the room and 
damaging furniture, and more horribly, hitting, slapping, and even choking his 
wife a couple of times a week during his sleep \citep{24}.

\section{The visceral theory of sleep and mental disorders}
In the previous section we described some transient sleep disorders caused
by inappropriate gating of external and visceral information streams in the 
brain. But what happens if for some reason some gate elements are not properly
functioning permanently? 

The striking similarity between psychotic phenomena during REM behavior 
disorder and symptomatics of schizophrenia lead to a hypothesis \citep{26,27}
that schizophrenia could be a REM disorder, a kind of permanent dream attack 
while awake, a ``waking reality processed through the dreaming brain'' 
\citep{26}.

The visceral theory of sleep prompts to suggest a stronger hypothesis: 
mental disorders are caused by failures of gating elements of the brain 
information processing system because of which visceral information is 
permanently present in brain's ``universal processor'' and during wakefulness
is interpreted by the brain as coming from external sources. So to say,
psychiatric illness is a condition when a dreaming reality caused by
unblocked visceral signals is processed as real by the waking brain. 

Wrong switching of the visceral information stream is harmful not only because
it can lead to a distortion of perceived reality and as a result to the brain's
irrational response. It seems plausible to imagine that a situation when 
a gate (or some gates) on the way of visceral information stream remains open 
during wakefulness is stressful for the brain's information processing system 
because the majority of the visceral signals will be interpreted as a noise. 
A continuous struggle with this ``noise'' can exhaust brain's adaptive 
mechanisms and in a long run lead to dysregulation and breakdown of  
mental functions and abilities not directly connected with the faulted gates. 
	
The well known fact that sleep and circadian rhythm disorders accompany almost 
every psychiatric illness \citep{28,29,30} can be considered as an indirect 
confirmation that visceral theory of sleep can indeed shed some light on mental
disorders. For example, a recent study \citep{30A} of the literature on sleep 
disorders and schizophrenia reveals an overwhelming evidence that sleep 
abnormalities, such as insomnia, obstructive sleep apnea, restless leg 
syndrome, are pervasive in schizophrenia patients. On the other side, the
findings of \citep{30B} firmly indicate that a disruption of  circadian 
rhythmicity and sleep is a core feature of mood disorders.

The dominant model for sleep-wake cycle regulation today is the two-process 
model of Borb\'{e}ly \citep{31A,31}. According to this 
model two processes,  a homeostatic process and a clock-like process 
controlled by the circadian pacemaker, govern the sleep-wake cycle. 
Interestingly, Borb\'{e}ly and Wirz-Justice immediately gave an application of 
this model to mental disorders \citep{32}, and
multiple lines of evidence suggest that bipolar disorder can be 
considered to a large extent as a disorder of circadian rhythms and sleep-wake 
processes \citep{33,34}.

Circadian system utilizes a special non-visual photoreceptors, different from 
the usual rods and cones of the mammalian retina, to differentiate darkness 
from light and synchronize the biological clock with the natural day-light 
cycle. Although their existence was suspected by Clyde Keeler already in 1927,
such photoreceptors were discovered only recently \citep{35,36}. 

Circadian photoreceptors use a photopigment, most probably melanopsin, with
a peak sensitivity at approximately 480 nm (blue light). If dysregulation of 
the circadian rhythms indeed plays the central role in symptomatics of bipolar
disorder, it is expected that manipulating a blue light exposure of bipolar
patients may have therapeutic potential. Interestingly, preliminary studies 
indeed report mood stabilization and reduced manic behavior in bipolar 
patients who wear at night special glasses with amber lenses that block blue 
light from artificial sources such as electric lights, televisions, and 
computers \citep{37,38}. 

A particularly striking example of the effectiveness of light therapy (or 
rather more precisely, dark therapy) is described in \citep{39}. A bipolar 
patient cycled rapidly between depression and mania. No medications were used
for his treatment. Instead he was asked to remain at bed in complete darkness 
for 14 h each night. The response to this dark therapy was dramatic: 
notwithstanding severe and unremitting rapid cycling for several years prior
to the treatment, his mood stabilized within several weeks and he experienced 
a complete cessation of cycling. 

As is well known, Lithium is the classic mood stabilizer commonly used in the 
treatment of the bipolar disorder. It seems this therapeutic effect of Lithium
is due to its ability to alter circadian rhythms. As various studies indicate 
\citep{40}, Lithium acts directly on the molecular clocks, although its 
chronobiological effects are not completely understood.

Within the hypothesized connection between visceral theory of sleep and 
origins of mental disorders, the association between circadian rhythm 
disruptions and bipolar disorder is understood through the role circadian 
rhythm plays in the control of gating elements of brain's information 
processing system.

We can expect that the continues presence of visceral noise during wakefulness
can affect visual and other perceptions. The extreme manifestations of this are
hallucinations inherent of schizophrenia. However more delicate effects are
also not excluded if the level of noise is not very high. In this respect 
binocular rivalry presents an interesting test ground for such putative 
influence.

When different images are presented to each eye simultaneously, instead of a 
superposition of the two images, perception alternates spontaneously between 
competing monocular views every few seconds. This phenomenon is called 
binocular rivalry and some research indicate that noise might be a crucial 
force in rivalry, frequently dominating the deterministic forces \citep{41,42}.

In light of our hypotheses that the visceral noise plays a fundamental role
in mental disorders, we expect that the perceptual rivalry will be affected
in patients with psychiatric illnesses. Some evidence shows that this is 
indeed the case. It was found \citep{43,44} that in bipolar patients 
alternation rates were slower than in normal controls even during remission, 
while the rates during depressive episodes were significantly slower than 
during remission. 

In some studies schizophrenia and major depression patients don't show the 
altered rivalry rates compared to the control group \citep{45}. The transition 
between the two alternate percepts in binocular rivalry is not always clean, 
however, and a mixing of the two images may be seen. It was found \citep{46} 
that  patients with schizophrenia showed a trend toward less mixed perceptions 
than controls. Besides, other studies found consistently slower binocular 
rivalry rates in participants with schizophrenia with the conclusion that 
abnormally-slow visual processing may be a feature of psychosis in general 
rather than a feature specific to bipolar disorder \citep{47}. 

Motion-induced blindness is another interesting oscillatory phenomenon in 
visual perception in which small but salient visual targets disappear 
intermittently from the visual awareness when surrounded by a global moving 
pattern. Sometimes it is considered just as a form of perceptual rivalry
because its temporal oscillation pattern of disappearance and reappearance is
highly correlated with the pattern reported for binocular rivalry in the same
individual \citep{48}. Interestingly but consistently with what was found in
binocular rivalry,  significantly lower rates of the motion-induced blindness
were found in  schizophrenia-spectrum disorder participants than in the control
group \citep{49}.

Somewhat paradoxically, schizophrenia patients show diminished susceptibility
to optical illusions. For example,  to healthy participants a hollow mask will 
look convex,  perceived as a normal face even when they are really seeing the 
concave side, whereas  patients with schizophrenia do not perceive the 
hollow-mask illusion \citep{50,51}. Recent brain-imaging and electrophysiology 
studies   are compatible with the idea that individuals with schizophrenia and 
controls use different perceptual strategies to minimize errors in visual 
awareness \citep{52}. Within the framework of our hypothesis, this change of 
perceptual strategy may be attributed to the presence of significant visceral 
noise in visual channel. An interesting fact that, similarly with the patient 
group, sleep-deprived medical staff show a remarkable impairment of binocular 
depth inversion \citep{50} indirectly supports this supposition in light of 
the visceral theory of sleep. 

The Necker cube is a classic example of bistable perception. As  perceptual 
reversals of ambiguous figures have several features in common with the
binocular rivalry \citep{53}, it is not surprising that the rate of perceptual
alternation of the Necker cube was reported to be slower in patients with
bipolar disorder (but not in patients with schizophrenia) \citep{54}. 
Interestingly, neural network simulation of Necker cube perception suggests
a view that manic states arise from excessive levels of cortical noise that 
destabilize neural representations \citep{55}. In case of mild mania, this 
noise might be even beneficial because it increases the speed in which the 
perceptual system abandons one particular scheme of organization of perceptual
information in favor of reassembling these same elements into a new meaningful
scheme. As a result creative thinking is facilitated due to this ability to 
parse elements of experience in a different and unexpected way \citep{55}. 
There exists an ample evidence of the link between creativity and bipolar 
disorder \citep{56,57,58,59}. According to Jamison \citep{58}, biographical 
data suggests that bipolar disorder may have affected, among others, Hemingway,
Faulkner, Fitzgerald, Dickens, O'Neill, Woolf, Handel, Ives, Rachmaninoff, 
Tchaikovsky, Keats, Gauguin, O'Keefe, Munch, Pollock. However still many 
questions do remain regarding this mysterious connection \citep{59}. 

\section{Concluding remarks}
The main idea of visceral theory of sleep is that the processing of external
and internal information streams by brain are largely separated in time. 
The same cortical neurons that process exteroceptive information in 
wakefulness switch during sleep to the processing of the interoceptive 
information. Thus during sleep the brain plays a dual role: from one side it 
is involved in the scanning of all life supporting systems, and processing of 
the relevant visceral information, and from the other side the brain itself is 
a paramount visceral organ also requiring control and management of its state.
 
However such a scheme of organization of the main information flows in the 
brain logically implies a possibility of errors and desynchronization in 
switching of the external and internal information streams by various gating 
elements of the brain information processing architecture. Anticipated
synchronization errors allow us to explain various transient pathological 
states connected with sleep, as indicated above. But if gate elements are 
not properly functioning constantly more serious consequences may appear. 

We hypothesize in this article that the primary cause of mental illnesses is 
just such a permanent fault in switching of the external and internal 
information streams during sleep-wake cycle as a result of which visceral 
signals are permanently present in the cerebral cortex biasing the normal 
information processing.
 
Of course, without a detailed knowledge of the mechanisms of interactions of 
the external and internal  information flows in the body our hypothesis has 
little practical significance. However, notwithstanding the mentioned 
drawback, we hope that the hypothesized connection between visceral theory of 
sleep and origins of mental disorders will stimulate a fresh look and research 
on the connections between the organization principles of the brain 
information processing and mental illnesses.

\section*{Acknowledgements}
We thank L.~V.~Tsibizov for drawing our attention to the visceral theory of
sleep, as well as I.~N.~Pigarev and M.~Lebedev for their support.


\begin{thebibliography}{10}

\bibitem{1}
Grande I,  Berk M, Birmaher B, Vieta E. Bipolar disorder.
{\it Lancet} 2016; {\bf 387}: 1562-1572.

\bibitem{2}
Baker J.A. Bipolar disorders: an overview of current literature.
{\it Journal of Psychiatric and Mental Health Nursing} 2001; {\bf 8}: 437-441.

\bibitem{3}
Belmaker R.H. Bipolar Disorder.
{\it The new England Journal of Medicine} 2004; {\bf 351}: 476-486.

\bibitem{4}
Maletic V, Raison C. Integrated neurobiology of bipolar disorder.
{\it Frontires in Psychiatry} 2014; {\bf 5}: 98.

\bibitem{5}
Phelps J.
Brain Differences in Bipolar Disorder,
\newline {\small
http://psycheducation.org/the-biologic-basis-of-bipolar-disorder/1035-2/
}

\bibitem{6}
Peng D, Jiang K. Comorbid bipolar disorder and obsessive-compulsive disorder.
{\it Shanghai Archives of Psychiatry} 2015; {\bf 27}: 246-248.

\bibitem{7}
Muneer A. Bipolar Disorder: Role of Inflammation and the Development of 
Disease Biomarkers.
{\it Psychiatry Investigation} 2016; {\bf 13}: 18-33.

\bibitem{8}
Maclean P.D. Psychosomatic Disease and the "Visceral Brain": Recent 
Developments Bearing on the Papez Theory of Emotion.
{\it Psychosomatic Medicine} 1949; {\bf 11}: 338-353.

\bibitem{9}
Dalgleish T. The emotional brain.
{\it Nature Reviews Neuroscience} 2004; {\bf 5}: 583-589.

\bibitem{10}
Wimmer R.D, Schmitt L.I, Davidson T.J, Nakajima M, Deisseroth K, Halassa M.M.
Thalamic control of sensory selection in divided attention.
{\it Nature} 2015; {\bf 526}: 705-709.

\bibitem{11}
Frank M.G. The mystery of sleep function: current perspectives and future 
directions. 
{\it Reviews in the Neurosciences} 2006; {\bf 17}: 375-392.

\bibitem{12}
Assefa S.Z, Diaz-Abad M, Wickwire E.M, Scharf S.M. The Functions of Sleep.
{\it AIMS Neuroscience} 2015; {\bf 2(3)}: 155-171.

\bibitem{13}
Rial R.V, Nicolau M.C, Gamund\'{i} A, Aka\^{a}rir M, Aparicio S, Garau C, 
Tejada S, Roca C, Gen\'{e} L, Moranta D, Esteban S. The trivial function of 
sleep.
{\it Sleep Medicine Reviews} 2007; {\bf 11}: 311-325.

\bibitem{14}
Rattenborg N.C, Lesku J.A, Martinez-Gonzalez D, Lima S.L. The non-trivial 
functions of sleep.
{\it Sleep Medicine Reviews} 2007; {\bf 11}: 405-409.

\bibitem{15}
Pigarev I.N. Neurons of visual cortex respond to visceral stimulation during 
slow wave sleep.
{\it Neuroscience} 1994; {\bf 62}: 1237-1243.

\bibitem{16}
Pigarev I.N. The Visceral Theory of Sleep.
{\it Neuroscience and Behavioral Physiology} 2014; {\bf  44}: 421-434.

\bibitem{17}
Pigarev I.N, Pigareva M.L. The state of sleep and the current brain paradigm.
{\it Frontiers in Systems Neuroscience} 2015; {\bf 9}: 139.

\bibitem{18}
Pigarev I.N, Pigareva M.L. Sleep and the Control of Visceral Functions.
{\it Neuroscience and Behavioral Physiology} 2012; {\bf 42}: 948-956.

\bibitem{19}
Pigarev I.N, Pigareva M.L. Partial sleep in the context of augmentation of 
brain function.
{\it Frontiers in Systems Neuroscience} 2014; {\bf 8}: 75.

\bibitem{20}
Gy\"{o}rgy \'{A}. {\it Visceral Perception: Understanding Internal Cognition.}
New York: Plenum Press, 1998. 

\bibitem{21}
Rama A.N, Kushida C.A. Restless legs syndrome and periodic limb movement 
disorder.
{\it Medical Clinics of North America} 2004; {\bf 88}: 653-667.

\bibitem{22}
Basu P.P, Shah N.J, Krishnaswamy N, Pacana T. Prevalence of restless legs 
syndrome in patients with irritable bowel syndrome.
{\it World Journal of Gastroenterology} 2011; {\bf 17(39)}: 4404-4407.

\bibitem{23}
Umanath S, Sarezky D, Finger S. Sleepwalking through history: me\-dicine, 
arts, and courts of law.
{\it Journal of the History of the Neurosciences} 2011; {\bf 20}: 253-276.

\bibitem{24}
Moorcroft W.H. {\it Understanding Sleep and Dreaming.}
New York: Sprin\-ger Science \& Business Media, 2013.

\bibitem{25}
Schenck C.H, Bundlie S.R, Ettinger M.G, Mahowald M.W.
Chronic behavioral disorders of human REM sleep: a new category of parasomnia.
{\it Sleep} 1986; {\bf 9(2)}: 293-308.

\bibitem{26}
Griffin J, Tyrrell I. {\it Dreaming reality: How dreaming keeps us sane
or can drive us mad.} Chelvington: HG Publishing, 2006.

\bibitem{27}
Mannaa M, Walker L. Is Schizophrenia a REM Disorder?
{\it Academic Journal of Pediatrics and Neonatology} 2017; {\bf 4(2)}: 555690. 

\bibitem{28}
Benca R.M, Obermeyer W.H, Thisted R.A, Gillin J.C. 
Sleep and psychiatric disorders. A meta-analysis.
{\it Archives of general psychiatry} 1992; {\bf 49(8)}: 651-668. 

\bibitem{29}
Krystal A.D. Psychiatric disorders and sleep.
{\it Neurologic Clinics}  2012; {\bf 30(4)}: 1389-1413.

\bibitem{30}
Asarnow L.D, Soehner A.M, Harvey A.G.
Circadian Rhythms and Psychiatric Illness.
{\it Current Opinion in Psychiatry} 2013; {\bf 26(6)}: 566-571.

\bibitem{30A}
Kaskie R.E, Graziano B, Ferrarelli F.
Schizophrenia and sleep disorders: links, risks, and management challenges.
{\it Nature and Science of Sleep} 2017; {\bf 9}: 227-239.

\bibitem{30B}
Lyall L.M, Wyse C.A, Graham N, Ferguson A, Lyall D.M, Cullen B, 
Celis Morales C.A, Biello S.M, Mackay D, Ward J, Strawbridge R.J, 
Gill J.M.R, Bailey M.E.S, Pell J.P, Smith D.J.
Association of disrupted circadian rhythmicity with mood disorders, subjective
wellbeing, and cognitive function: a cross-sectional study of 91 105 
participants from the UK Biobank.
{\it The Lancet Psychiatry} 2018; {\bf 5(2)}: 507-514

\bibitem{31A}
Borb\'{e}ly A.A. A two process model of sleep regulation.
{\it Human Neurobiology} 1982; {bf 1}: 195-204.

\bibitem{31}
Borb\'{e}ly A.A,  Daan S, Wirz-Justice A, Deboer T.
The two-process model of sleep regulation: a reappraisal.
{\it Journal of Sleep Research} 2016; {\bf 25(2)}: 131-143.

\bibitem{32}
Borb\'{e}ly A.A, Wirz-Justice A. Sleep, sleep deprivation and depression. 
A hypothesis derived from a model of sleep regulation.
{\it Human neurobiology} 1982; {\bf 1(3)}: 205-210.

\bibitem{33}
Soreca I. Circadian rhythms and sleep in bipolar disorder: implications for 
pathophysiology and treatment.
{\it  Current Opinion in Psychiatry} 2014; {\bf 27(6)}: 467-471.

\bibitem{34} 
Murray G, Harvey A. Circadian rhythms and sleep in bipolar disorder.
{\it Bipolar Disorders} 2010; {\bf 12}: 459-472.

\bibitem{35}
Van Gelder R.N, Buhr E.D. Ocular Photoreception for Circadian Rhythm 
Entrainment in Mammals.
{\it Annual Review of Vision Science} 2016; {\bf 2}: 153-169.

\bibitem{36}
Berson D.M. Strange vision: ganglion cells as circadian photoreceptors.
{\it Trends in Neurosciences} 2003; {\bf 26(6)}: 314-319.

\bibitem{37}
Phelps J. Dark therapy for bipolar disorder using amber lenses for blue light 
blockade.
{\it Medical Hypotheses} 2008; {\bf 70(2)}: 224-229.

\bibitem{38}
Henriksen T.E, Skrede S, Fasmer O.B, Hamre B, Gr{\o}nli J, Lund A.
Blocking blue light during mania --- markedly increased regularity of sleep 
and rapid improvement of symptoms: a case report.
{\it Bipolar Disorder} 2014; {\bf 16(8)}: 894-898.

\bibitem{39}
Wehr T.A, Turner E.H, Shimada J.M, Lowe C.H, Barker C, Leibenluft E. 
Treatment of rapidly cycling bipolar patient by using extended bed rest and 
darkness to stabilize the timing and duration of sleep.
{\it  Biological Psychiatry}  1998; {\bf 43}: 822-828.

\bibitem{40}
Moreira J, Geoffroy P.A. Lithium and bipolar disorder: Impacts from molecular 
to behavioural circadian rhythms.
{\it Chronobiology International} 2016; {\bf 33(4)}: 351-373.

\bibitem{41}
Brascamp J.W, van Ee R, Noest A.J, Jacobs R.H.A.H, van den Berg A.V.
The time course of binocular rivalry reveals a fundamental role of noise.
{\it Journal of Vision} 2006; {\bf 6}: 1244-1256.

\bibitem{42}
Roumani D, Moutoussis K. Binocular rivalry alternations and their relation 
to visual adaptation.
{\it Frontiers in Human Neuroscience} 2012; {\bf 6}: 35.

\bibitem{43}
Pettigrew J.D, Miller S.M. A 'sticky' interhemispheric switch in bipolar 
disorder?
{\it Proceedings of the Royal Society of London, Series B, Biological 
Sciences} 1998; {\bf 265}: 2141-2148.

\bibitem{44}
Jia T, Ye X, Wei Q, Xie W, Cai C, Mu J, Dong Y, Hu P, Hu X, Tian Y, Wang K.
Difference in the binocular rivalry rate between depressive episodes and 
remission.
{\it Physiology \& Behavior} 2015; {\bf 151}: 272-278.  

\bibitem{45}
Miller S.M, Gynther B.D, Heslop K.R, Liu G.B, Mitchell P.B, Ngo T.T, 
Pettigrew J.D, Geffen L.B. Slow binocular rivalry in bipolar disorder.
{\it Psychological Medicine} 2003; {\bf 33}: 683-692.

\bibitem{46}
Stanley J, Park S, Blake R, Carter O. Binocular rivalry dynamics and mixed 
percept in schizophrenia.
{\it Frontiers in Human Neuroscience} 2015; Conference Abstract: XII 
International Conference on Cognitive Neuroscience (ICON-XII).

\bibitem{47}
Heslop K.R. Binocular rivalry and visuospatial ability in individuals
with schizophrenia. PhD thesis, Queensland University of Technology. 2012.

\bibitem{48}
Carter O.L, Pettigrew J.D. A common oscillator for perceptual rivalries?
{\it Perception} 2003; {\bf 32}: 295-305.

\bibitem{49}
Tschacher W, Schuler D, Junghan U. Reduced perception of the motion-induced 
blindness illusion in schizophrenia.
{\it Schizophrenia Research} 2006; {\bf 81}: 261-267.

\bibitem{50}
Schneider U, Leweke F.M, Sternemann U, Weber M.M, Emrich H.M.
Visual 3D illusion: a systems-theoretical approach to psychosis.
{\it European Archives of Psychiatry and Clinical Neuroscience} 1996;
{\bf 246}: 256-260.

\bibitem{51}
Dima D, Roiser J.P, Dietrich D.E, Bonnemann C, Lanfermann H, Emrich H.M, 
Dillo W. Understanding why patients with schizophrenia do not perceive the 
hollow-mask illusion using dynamic causal modelling.
{\it NeuroImage} 2009; {\bf 46}: 1180-1186.

\bibitem{52}
Notredame C.-E, Pins D, Deneve S, Jardri R. What visual illusions teach us 
about schizophrenia.
{\it Frontiers in Integrative Neuroscience} 2014; {\bf 8}: 63. 

\bibitem{53}
Kornmeier J, Bach M. The Necker cube --- an ambiguous figure disambiguated in 
early visual processing.
{\it Vision Research} 2005; {\bf 8}:  955-960.

\bibitem{54}
Hunt J.M, Guilford J.P. Fluctuation of an ambiguous figure in dementia praecox 
and in manic-depressive patients. 
{\it The Journal of Abnormal and Social Psychology} 1933; {\bf 27(4)}: 443-452.

\bibitem{55}
Hoffman R.E, Quinlan D.M, Mazure C.M, McGlashan T.M. Cortical instability and 
the mechanism of mania: a neural network simulation and perceptual test.
{\it Biological Psychiatry} 2001; {\bf 49}: 500-509.

\bibitem{56}
Andreasen N.C, Glick I.D. Bipolar affective disorder and creativity: 
Implications and clinical management.
{\it Comprehensive Psychiatry} 1988; {\bf 29}: 207-217.

\bibitem{57}
Jamison K.R. Mood disorders and patterns of creativity in British writers and 
artists.
{\it Psychiatry: Interpersonal and Biological Processes} 1989; {\bf 52(2)}:
125-134.

\bibitem{58}
Jamison K.R. {\it Touched with fire: Manic-Depressive Illness and the Artistic 
Temperament.} New York: Free Press, 1993.

\bibitem{59}
Johnson S.L, Murray G, Fredrickson B, Youngstrom E.A, Hinshaw S, Bass J.M, 
Deckersbach T, Schooler J, Salloum I. Creativity and bipolar disorder: touched 
by fire or burning with questions?
{\it Clinical Psychology Review} 2012; {\bf 32}: 1-12.

\end{thebibliography}


\end{document}